\documentclass{article}
\usepackage{times}
\usepackage[pdfmark,colorlinks]{hyperref}
\usepackage{amsfonts}
\begin{document}
\title{Magnetic Backgrounds\\ from \\ Generalised Complex Manifolds}
\author{Jos\'e M. Isidro\\
Instituto de F\'{\i}sica Corpuscular (CSIC--UVEG)\\
Apartado de Correos 22085, Valencia 46071, Spain\\
{\tt jmisidro@ific.uv.es}}

\maketitle

\begin{abstract}

\noindent
The magnetic backgrounds that physically give rise to spacetime noncommutativity are generally treated using noncommutative geometry. In this article we prove that also the theory of generalised complex manifolds contains the necessary elements to generate $B$--fields geometrically. As an example, the Poisson brackets of the Landau model (electric charges on a plane subject to an external, perperdicularly applied magnetic field) are rederived using the techniques of generalised complex manifolds.

\end{abstract}

\tableofcontents

\section{Introduction}\label{ramalloguarrolavatelospies}

The groundbreaking papers of ref. \cite{CDS} concerning the appearance of noncommutative geometry in string theory with a nonvanishing $B$--field triggered off an avalanche of activity (for reviews see, {\it e.g.}, ref. \cite{SZABO}).  Although the subject of physics in the presence of a magnetic background became fashionable around the turn of the century, the notion of noncommuting space variables actually arose decades before. Thus, textbooks on quantum mechanics often treat the Landau levels of a nonrelativistic charge moving on the $x,y$ plane traversed by a uniform magnetic field ${\bf B}=(0,0,\vert {\bf B}\vert)$, a problem where noncommuting momenta arise naturally \cite{LANDAU}. One has the quantum commutator
\begin{equation}
\left[P_x,P_y\right]={\rm i}\hbar\frac{e\vert {\bf B}\vert}{c}.
\label{ramallocasposo}
\end{equation}

Ever since Einstein's times one is used to the idea that geometry has something to say in physics, often quite a lot. The question immediately arises, {\it what is the geometry underlying the $B$--field?} One possible answer is to resort to the theory of $C^{\star}$--algebras and noncommutative geometry \cite{NCG}, as done in ref. \cite{CDS}. An alternative approach is to look for usual ({\it i.e.}, commutative) manifolds endowed with some additional structure. A technical description of the $B$--field can be given in terms of connections on gerbes \cite{HITCHIN}. Without using gerbes, the necessary geometry cannot just be based on invariance under diffeomorphisms of some kind---we will see that it must go beyond diffeomorphic invariance. The geometry of the $B$--field, while making use of diffeomorphisms, relies critically on additional tools. Thus, when turning on a magnetic field, the Poisson brackets $\left\{p_x,p_y\right\}=0$ cease to vanish if one continues to use the same Darboux coordinates used in the absence of the $B$--field. Therefore this new geometry escapes Klein's {\it Erlanger Programm}\/ in the sense that it cannot be characterised simply as being invariant under some group of transformations. 

Recent advances in the theory of generalised complex manifolds (first appeared in \cite{NIGEL} and developed more fully in \cite{GUALTIERI}) have paved the way for a deeper understanding of the mathematics underlying a lot of interesting contemporary physics; some works along these lines are collected in ref. \cite{TODOS}. In this article we prove that the theory of generalised complex manifolds as developed in ref. \cite{GUALTIERI} contains the necessary elements to geometrically generate the magnetic backgrounds that physically give rise to spacetime noncommutativity. As indicated above, our viewpoint must be regarded as complementing that of noncommutative geometry. For simplicity, rather than considering a string or brane--theoretic setup, we will restrict ourselves to a mechanical problem (classical and quantum) with a finite number of degrees of freedom. This is also convenient since, to the best of our knowledge, infinite--dimensional, generalised complex manifolds have not yet been studied. Presumably an infinite--dimensional extension of the formalism of ref. \cite{GUALTIERI} must exist, that will enlarge the theory of infinite--dimensional complex manifolds as presented, {\it e.g.}, in ref. \cite{NACHBIN}.

This article is organised as follows. Section \ref{barbobketedenporkulomarikondemierda} sets the scene with some well--known material on symplectic and complex geometry. A crash course in generalised complex manifolds is quickly presented in section \ref{pepebarboneatshit}, with an emphasis on the aspects that are needed for our purposes. Mechanics meets generalised complex geometry in section \ref{javiermasmariquitas}, where contact is made between these two disciplines. All the preceding geometry finds its physical application when a magnetic field is turned on, as in section \ref{ramayoketepartaunrayo}. Conclusions are finally presented in section \ref{javiermasimbecildelculo}.

\section{Some preliminaries}\label{barbobketedenporkulomarikondemierda}

The classical dynamics of a finite number $n$ of degrees of freedom is best described in terms of a classical phase space ${\cal C}$. 
The latter is real $2n$--dimensional admitting a symplectic structure with a symplectic form $\omega$. 
Consider the tangent and cotangent bundles to classical phase space, $T{\cal C}$ and $T^*{\cal C}$. 
The symplectic structure can be viewed as an isomorphism $\omega_x$ between the tangent and the cotangent fibres over each point 
$x\in {\cal C}$,
\begin{equation} 
\omega_x\colon T_x{\cal C}\longrightarrow T^*_x{\cal C},
\label{javiermasnecesitasvaselinaparaelkulo}
\end{equation}
satisfying 
\begin{equation}
\omega_x^t=-\omega_x
\label{javiermasmelasopla}
\end{equation}
for all $x\in {\cal C}$. The integrability condition ${\rm d}\omega=0$ must be satisfied \cite{ARNOLD}. In Darboux coordinates $q^j,p_j$ around $x\in{\cal C}$ we have 
\begin{equation}
\omega_x=\sum_{j=1}^n{\rm d}q^j\wedge{\rm d}p_j.
\label{barbonketefollenmarikon}
\end{equation}

Quantum--mechanically one is limited by Heisenberg's uncertainty principle, so the simultaneous specification of ${\rm d}q^j$ and ${\rm d}p_j$ 
in eqn. (\ref{barbonketefollenmarikon}) has a lower bound given by $\hbar/2$. From a geometrical perspective, 
quantum mechanics abandons symplectic geometry. Instead the holomorphic tangent bundle appears naturally as the bearer of quantum--mechanical 
information about the system. Two categories arise here: holomorphic objects and tangent objects. The quantum theory requires both of them.

That the category of complex manifolds arises naturally in quantum mechanics is best appreciated in the theory of coherent states \cite{PERELOMOV}. There is also a simple heurisitic argument in favour of holomorphic objects as appropriate for carrying quantum mechanical information. Namely, holomorphic objects naturally respect the limitations imposed by Heisenberg's principle because, roughly speaking, they depend on $z^j=(q^j+{\rm i}p_j)/\sqrt{2}$ but not on $\bar z^j=(q^j-{\rm i}p_j)/\sqrt{2}$. In this way the transformation from Darboux coordinates $q^j, p_j$ to holomorphic coordinates $z^j$ 
cannot be inverted, since inverting it would require using also the $\bar z^j$, thus spoiling holomorphicity. In other words, the theory expressed 
in terms of holomorphic coordinates $z^j$ only contains half as much information as the theory expressed in terms of Darboux coordinates 
$q^j, p_j$, and Heisenberg's principle is respected. Equivalently we may state that the passage from classical to quantum mechanics implies a certain loss of information, 
which is implemented mathematically through complexification of classical phase space.

Quantisation, however, does not stop at complexification. Once within the holomorphic category, we will further argue in favour of holomorphic {\it tangency}\/ as being quantum--mechanical in nature. For the moment let us recall that a complex structure $J$ on ${\cal C}$ is an endomorphism of the tangent fibre over each point $x\in{\cal C}$
\begin{equation}
J_x\colon T_x{\cal C}\longrightarrow T_x{\cal C}
\label{ramallotehanabiertolaspataspordetras}
\end{equation}
satisfying 
\begin{equation}
J_x^2=-{\bf 1}
\label{barbonketefollen}
\end{equation}
for all $x\in {\cal C}$, as well as the Newlander--Nirenberg integrability condition that the Nijenhuis tensor $N$  vanish identically \cite{KOBAYASHI}.
The latter is defined on holomorphic tangent vectors $Z$, $W$ on ${\cal C}$ in terms of the Lie brackets $[\cdot\,,\cdot]$ of vector fields as
\begin{equation}
N(Z,W):=[Z,W]-[JZ,JW]+J[JZ,W]-J[Z,JW].
\label{barbonkabron}
\end{equation}
So the definition of a complex structure requires the notion of tangency. 

Let the classical Darboux coordinates $q^j$, $p_j$ quantise to the quantum observables  $Q^j$, $P_j$. In the quantum theory, commutators arise as a natural composition law for operators. Commutators satisfy the properties of Lie brackets, which is the natural operation defined on tangent vectors. Hence we can think of the holomorphic tangent bundle $T_{(1,0)}{\cal C}$ (where $T_{(1,0)}{\cal C}\oplus T_{(0,1)}{\cal C}=T{\cal C}\otimes\mathbb{C}$) as being quantum--mechanical in nature. By contrast, the cotangent bundle $T^*{\cal C}$ was seen to be the relevant object in classical mechanics. We conclude that the quantum theory is best expressed in terms of {\it holomorphic, tangent}\/ objects.

\section{Generalised complex manifolds: the basics}\label{pepebarboneatshit}

We follow ref. \cite{GUALTIERI} closely, omitting geometrical technicalities for brevity. 
Thus, {\it e.g.}, we will illustrate our conclusions in local coordinates around a point $x\in {\cal C}$, forgetting about global issues 
that can be taken care of by the appropriate integrability conditions. Let our mechanics have $n$ degrees of freedom. Then the total space of the bundle $T{\cal C}\oplus T^*{\cal C}$ is real $6n$--dimensional: $2n$ dimensions for the base, $4n$ for the fibre.

The space $T_x{\cal C}\oplus T^*_x{\cal C}$ carries the inner product 
\begin{equation}
\langle X+\xi,Y+\eta\rangle:=\frac{1}{2}\left(\xi(Y)+\eta(X)\right),
\label{linprod}
\end{equation}
where $X,Y\in T_x{\cal C}$ and $\xi, \eta\in T_x^*{\cal C}$. This inner product has the signature $(2n,2n)$.
The group $SO(2n,2n)$ acts on $T_x{\cal C}\oplus T^*_x{\cal C}$. Its Lie algebra $so(2n,2n)$ decomposes as
\begin{equation}
\left(\begin{array}{cc}
A&\beta\\
B&-A^t
\end{array}\right),
\label{barbonketefollenporkulo}
\end{equation}
where $A\in{\rm End}(T_x{\cal C})$, its transpose $A^t\in{\rm End}(T^*_x{\cal C})$ and
\begin{equation}
B:T_x{\cal C}\longrightarrow T_x^*{\cal C}, \qquad \beta:T_x^*{\cal C}\longrightarrow T_x{\cal C}
\label{ketefollenbarbon}
\end{equation}
are skew, {\it i.e.}, $B^t=-B$, $\beta^t=-\beta$. We view $B$ as a 2--form in $\Lambda^2T_x^*{\cal C}$ via $B(X)=i_XB$. Taking $A=0=\beta$ and exponentiating,
\begin{equation}
{\rm exp}\left(\begin{array}{cc}
0&0\\
B&0
\end{array}\right)=\left(\begin{array}{cc}
{\bf 1}&0\\
B&{\bf 1}
\end{array}\right),
\label{barbonazocabronazo}
\end{equation}
we obtain the orthogonal transformation
\begin{equation}
X+\xi\longrightarrow X+\xi+i_XB.
\label{barbonazohijoputazo}
\end{equation}
These transformations are important in what follows. They are called {\it B--transformations}.

A {\it generalised complex structure}\/ over ${\cal C}$, denoted ${\cal J}$, is an endomorphism of the fibre over each $x\in {\cal C}$,
\begin{equation}
{\cal J}_x\colon T_x{\cal C}\oplus T^*_x{\cal C}\longrightarrow T_x{\cal C}\oplus T_x^*{\cal C},
\label{ramallotienesunagujeroprofundoenelkulo}
\end{equation}
satisfying the following three conditions.  First, for all $x\in {\cal C}$ one has 
\begin{equation}
{\cal J}_x^2=-{\bf 1}.
\label{ramalloketehostien}
\end{equation}
Second, for all $x\in {\cal C}$ one has 
\begin{equation}
{\cal J}_x^t=-{\cal J}_x.
\label{barboncortatelospelosketienescaspa}
\end{equation}
Third, the Courant integrability condition must hold \cite{GUALTIERI}; in what follows we will assume that this latter condition is always satisfied. It should be realised that generalised complex geometry involves an object ${\cal J}$ that is simultaneously complex (eqn. (\ref{ramalloketehostien})) and symplectic (eqn. (\ref{barboncortatelospelosketienescaspa})). 

Suppose that ${\cal J}$ at $x\in {\cal C}$ is given by
\begin{equation}
{\cal J}_{\omega_x}=\left(\begin{array}{cc}
0&-\omega_x^{-1}\\
\omega_x&0\end{array}\right),
\label{ramallocomprateunchampuanticaspa}
\end{equation}
$\omega$ being a symplectic form as in eqn. (\ref{javiermasnecesitasvaselinaparaelkulo}). This ${\cal J}_{\omega}$ defines a generalised complex structure 
{\it of symplectic type}. At the other end we have that
\begin{equation}
{\cal J}_{J_x}=\left(\begin{array}{cc}
-J_x&0\\
0&J^t_x\end{array}\right),
\label{ramallotieneslospelosllenosdecaspa}
\end{equation}
$J$ being a complex structure as in eqn. (\ref{ramallotehanabiertolaspataspordetras}), defines a generalised complex structure {\it of complex type}. 

There exists a Darboux--like theorem describing the local form of a generalised complex structure in the neighbourhood of any regular point. Roughly speaking, any manifold endowed with a generalised complex structure splits {\it locally}\/ as the product of a complex manifold times a symplectic manifold. A more precise statement is as follows. A point $x\in{\cal C}$ is said {\it regular}\/ if the Poisson structure $\omega^{-1}$ has constant rank in a neighbourhood of $x$. Then any regular point in a generalised complex manifold has a neighbourhood which is equivalent, via a diffeomorphism and a $B$--transformation, to the product of an open set in $\mathbb{C}^k$ and an open set in $\mathbb{R}^{2n-2k}$, the latter endowed with its standard symplectic form. The nonnegative integer $k$ is called the {\it type}\/ of ${\cal J}$, $k=0$ and $k=n$ being the limiting cases examined in eqns. (\ref{ramallocomprateunchampuanticaspa}) and (\ref{ramallotieneslospelosllenosdecaspa}), respectively.

Next assume that ${\cal C}$ is a linear space. Then any generalised complex structure of type $k=0$ is the $B$--transform of a symplectic structure. This means that any generalised complex structure of type $k=0$ can be written as
$$
{\rm e}^{-B}{\cal J}_{\omega}{\rm e}^{B}=
\left(\begin{array}{cc}
{\bf 1}&0\\
-B&{\bf 1}\end{array}\right)
\left(\begin{array}{cc}
0&-\omega^{-1}\\
\omega &0\end{array}\right)
\left(\begin{array}{cc}
{\bf 1}&0\\
B&1\end{array}\right)
$$
\begin{equation}
=\left(\begin{array}{cc}
-\omega^{-1}B & -\omega^{-1}\\
\omega + B\omega^{-1}B & B\omega^{-1}\end{array}\right),
\label{ramalloestasllenodecaspa}
\end{equation}
for a certain 2--form $B$. Similarly any generalised complex structure of type $k=n$ over a linear manifold ${\cal C}$ is the $B$--transform of a complex structure,
$$
{\rm e}^{-B}{\cal J}_{J}{\rm e}^{B}=
\left(\begin{array}{cc}
{\bf 1}&0\\
-B&{\bf 1}\end{array}\right)
\left(\begin{array}{cc}
-J&0\\
0 & J^t\end{array}\right)
\left(\begin{array}{cc}
{\bf 1}&0\\
B&{\bf 1}\end{array}\right)
$$
\begin{equation}
=\left(\begin{array}{cc}
-J & 0\\
BJ+J^tB & J^t\end{array}\right).
\label{ramalloduchatekehuelesamierda}
\end{equation}
When ${\cal C}$ is an arbitrary smooth manifold, not necessarily a linear space, the previous statements hold essentially true, with some refinements required; see ref. \cite{GUALTIERI} for details.

\section{Mechanics meets generalised complex geometry}\label{javiermasmariquitas}

A metric of signature $(2n,2n)$ is readily manufactured with the mechanical elements at hand. Let an ordered basis for $T_x^*{\cal C}$ be spanned by ${\rm d}q^1,\ldots, {\rm d}q^n$, ${\rm d}p_1, \ldots, {\rm d}p_n$. Correspondingly, an ordered basis for $T_x{\cal C}\oplus T_x^*{\cal C}$ is spanned by $\partial_{q^1},\ldots, \partial_{q^n}$, $\partial_{p_1}, \ldots, \partial_{p_n}$,  ${\rm d}q^1,\ldots, {\rm d}q^n$, ${\rm d}p_1, \ldots, {\rm d}p_n$. Starting from the classical symplectic form $\omega$ in Darboux coordinates, its (block) matrix at $x\in{\cal C}$ is\footnote{Our conventions are such that $\omega=\frac{1}{2}\omega_{jk}{\rm d}x^j\wedge{\rm d}x^k$ and $\left\{f,g\right\}=\frac{1}{2}\pi^{jk}\partial_jf\partial_kg$.}
\begin{equation}
\omega=\left(\begin{array}{cc}
0&{\bf 1}_n\\
-{\bf 1}_n&0\end{array}\right).
\label{barbonketehostienporladron}
\end{equation}
The Poisson tensor $\pi$ is the inverse of the symplectic matrix $\omega$ \cite{ARNOLD},
\begin{equation}
\pi=\omega^{-1}.
\label{ramallomierda}
\end{equation}
In Darboux coordinates we have $\omega^{-1}=-\omega$ and $(-\omega)^2=-{\bf 1}$. Now ${\rm i}\hbar$ times classical Poisson brackets are quantum commutators. Hence the latter are represented by 
\begin{equation}
{\rm i}\hbar\left(\begin{array}{cc}
0&-{\bf 1}_n\\
{\bf 1}_n&0\end{array}\right).
\label{barbonketehostienpormarikon}
\end{equation}
Setting $\hbar=1$, the above squares to the identity. The direct sum of the squares of the matrices (\ref{barbonketehostienpormarikon}), (\ref{barbonketehostienporladron}) gives us the expression, in local coordinates, of a diagonal metric $G$ on $T{\cal C}\oplus T^*{\cal C}$
\begin{equation}
G=\left(\begin{array}{cc}
{\bf 1}_{2n}&0\\
0&-{\bf 1}_{2n}\end{array}\right).
\label{barboneatshit}
\end{equation}
Indeed, (\ref{barboneatshit}) is the diagonal form of the metric (\ref{linprod}). The Poisson brackets on classical phase space, plus the quantisation prescription that ${\rm i}\hbar$ times Poisson brackets become quantum commutators, automatically dictate that the negative--signature piece of the metric must be classical, while the piece with positive signature must be quantum.

The fact that a metric of the required signature can be constructed from the (classical and quantum) mechanical elements present in $T{\cal C}\oplus T^*{\cal C}$ indicates that our mechanical setup can make contact, in a natural way, with the geometry of generalised complex manifolds. Hence it makes sense to ask how mechanics transforms under $B$--transformations. This is analysed next.

\section{$B$--transformation of the classical Poisson brackets}\label{ramayoketepartaunrayo}

\subsection{Noncommuting momenta}\label{luisibanez,kalvoperokasposo}

When the symplectic form $\omega$ is expressed in Darboux coordinates as in eqns. (\ref{barbonketefollenmarikon}) and (\ref{barbonketehostienporladron}), the canonical Poisson brackets read
\begin{equation}
\left\{q^j,p_k\right\}=\delta^j_k, \qquad\left\{q^j, q^k\right\}=0=\left\{p_j,p_k\right\}.
\label{barbonketemetanuncilindro}
\end{equation}
Under a $B$--transformation, the lower left entry of ${\cal J}_{\omega}$ in  eqn. (\ref{ramalloestasllenodecaspa}) transforms as
\begin{equation}
\omega\longrightarrow\omega_B=\omega+B\omega^{-1}B.
\label{ramallokaka}
\end{equation}
The transformation (\ref{ramallokaka}) is reminiscent of a similar one in symplectic geometry, whereby a symplectic form $\tilde\omega$ transforms, under the action of a magnetic field $\tilde B$, as \cite{MR}
\begin{equation}
\tilde\omega\longrightarrow\tilde\omega_{\tilde B}=\tilde\omega-\tilde B.
\label{luisiabnezmekagoentuputasombra}
\end{equation}
Here $\tilde\omega_{\tilde B}$ qualifies as a new symplectic form whenever $\tilde B$ is a closed 2--form. Although analogous to (\ref{luisiabnezmekagoentuputasombra}), our transformation law (\ref{ramallokaka}) is quadratic in the magnetic field, instead of linear. Now $\omega_B$ in (\ref{ramallokaka}) continues to be closed and antisymmetric because $B$ itself is a closed 2--form. Hence the only property of $\omega$ that $\omega_B$ might eventually lose is nondegeneracy. However this possibility is ruled out by the results of ref. \cite{MR}, as applied to the closed 2--form 
\begin{equation}
\tilde B:=-B\omega^{-1}B, 
\label{barbonputonverbenero}
\end{equation}
and $\omega_B$ in (\ref{ramallokaka}) is nondegenerate.

Let us consider the generalised complex structure ${\cal J}_{\omega_B}$ of type $k=0$ defined by
\begin{equation}
{\cal J}_{\omega_B}=\left(\begin{array}{cc}
0&-\omega_B^{-1}\\
\omega_B&0\end{array}\right).
\label{lagmekagoentuputamadre}
\end{equation}
The $B$--transform $\pi_B$ of the Poisson tensor $\pi$
\begin{equation}
\pi_B:=\omega_B^{-1}=(\omega+B\omega^{-1}B)^{-1}
\label{putoramallo}
\end{equation}
can be computed order by order in powers of $B$, at least for $B$ sufficiently weak, as 
\begin{equation}
\pi_B=-\omega\sum_{j=0}^{\infty}(-1)^j(\omega B\omega B)^j.
\label{barbonchupameelpiton}
\end{equation}
The transformation laws (\ref{ramallokaka}), (\ref{putoramallo}) are nontensorial. In fact $B$--transformations are {\it not}\/ diffeomorphisms of ${\cal C}$, so it would be a surprise if $\omega$ and $\pi$ transformed tensorially under $B$. The Darboux coordinates that reduce $\omega$ to canonical form need no longer be Darboux coordinates for $\omega_B$. Thus generically the Poisson brackets $\left\{p_k,p_l\right\}_{\pi_B}$ will not vanish, which is interpreted physically as due to the existence of a background magnetic field. We will show that the $B$--transformed Poisson brackets of the initial canonical brackets (\ref{barbonketemetanuncilindro}) are those corresponding to the mechanical system one starts out with, but now submitted to an external magnetic field. However, rather than computing the inverse (\ref{putoramallo}) in the general case, we can proceed more simply as follows. 

To begin with let our phase space ${\cal C}$ be $\mathbb{R}^4$. We will show that the $B$--transformed Poisson brackets are those from which the quantum commutator (\ref{ramallocasposo}) derives. In fact this result follows from the theorem of ref. \cite{GUALTIERI} stated in eqn. (\ref{ramalloestasllenodecaspa}), but we can provide an alternative elementary proof, that will also be useful in what follows. In the ordered basis ${\rm d}x,{\rm d}y$, ${\rm d}p_x,{\rm d}p_y$, the matrix $\Pi^{jk}$ corresponding to the Poisson brackets 
\begin{equation}
\left\{p_x,p_y\right\}=\frac{e \vert{\bf B}\vert}{c}, \quad \left\{x,p_x\right\}=1,\quad \left\{y,p_y\right\}=1
\label{ramallocasposon}
\end{equation}
is
\begin{equation}
\Pi^{jk}=\left(\begin{array}{cccc}
0&0&1&0\\
0&0&0&1\\
-1&0&0&\kappa\\
0&-1&-\kappa&0
\end{array}\right), \qquad \kappa=\frac{e\vert {\bf B}\vert}{c}.
\label{ramayohijoputa}
\end{equation}
Its inverse $(\Pi^{jk})^{-1}$ is the matrix of the corresponding symplectic form  $\Omega_{jk}$,
\begin{equation}
\Omega_{jk}=\left(\begin{array}{cccc}
0&\kappa&-1&0\\
-\kappa&0&0&-1\\
1&0&0&0\\
0&1&0&0
\end{array}\right).
\label{ramayohijoputon}
\end{equation}
Our problem reduces to proving that an antisymmetric matrix $B$ exists, such that the corresponding $B$--transform of the symplectic form (\ref{barbonketehostienporladron}) equals the matrix $\Omega_{jk}$ of (\ref{ramayohijoputon}). The hurried reader can jump to the solution, eqn. (\ref{javiermasmarikas}). Since all our matrices are constant, ${\rm d}B=0$ will hold true.

Let the $B$--field be given by
\begin{equation}
B=\left(\begin{array}{cc}
M&-P^t\\
P&Q
\end{array}\right),\qquad M^t=-M, Q^t=-Q,
\label{ketefostienramallo}
\end{equation}
where $M,P,Q$ are $2\times 2$ matrices. Using eqns. (\ref{barbonketehostienporladron}), (\ref{ramallokaka}) we find
\begin{equation}
\omega_B=\left(\begin{array}{cc}
-P^tM-MP&{\bf 1}+(P^t)^2-MQ\\
-{\bf 1}+QM-P^2&-QP^t-PQ
\end{array}\right).
\label{ketefostienramallokabron}
\end{equation}
Imposing the equality of (\ref{ketefostienramallokabron}) and (\ref{ramayohijoputon}) gives the system of independent equations
\begin{equation}
(MP)^t-MP=\left(\begin{array}{cc}
0&\kappa\\
-\kappa&0
\end{array}\right),
\label{mekagonramallo}
\end{equation}
\begin{equation}
(PQ)^t-PQ=0,
\label{makagoentuputakararamayo}
\end{equation}
\begin{equation}
QM=P^2+2.
\label{luisibanezmekagontuputakalva}
\end{equation}
Denoting
\begin{equation}
MP:=R^t, \qquad PQ:=S^t,
\label{ramalloketejodanenprofundidad}
\end{equation}
eqns. (\ref{mekagonramallo}), (\ref{makagoentuputakararamayo}) give conditions on the antisymmetric projections of $R$ and $S$:
\begin{equation}
R-R^t=\left(\begin{array}{cc}
0&\kappa\\
-\kappa&0
\end{array}\right),
\label{mekagonramayo}
\end{equation}
\begin{equation}
S-S^t=0.
\label{ramalloketepartaunrayo}
\end{equation}
Thus $S$ must be symmetric, while $R$ cannot be symmetric. By eqns. (\ref{luisibanezmekagontuputakalva}), (\ref{ramalloketejodanenprofundidad}), 
\begin{equation}
P^4+2P^2-(RS)^t=0.
\label{pauli}
\end{equation}
Now by eqn.  (\ref{ramalloketejodanenprofundidad}),
\begin{equation}
Q=P^{-1}S^t, \qquad 
M=R^tP^{-1},
\label{pajaresimbecildelculo}
\end{equation}
ant the resulting $Q$, $M$ are required to be antisymmetric. Since $S$ must be symmetric, this requires that $P^t\neq P$ unless $S=0$.

In order to solve (\ref{mekagonramayo}), (\ref{ramalloketepartaunrayo}) and (\ref{pauli}) the most general {\it Ansatz} is
\begin{equation}
R=\frac{1}{2}\left(\begin{array}{cc}
a&\kappa\\
-\kappa& b
\end{array}\right),\quad
S=\left(\begin{array}{cc}
c&d\\
d&f
\end{array}\right), \quad
P=\left(\begin{array}{cc}
p&q\\
r&s
\end{array}\right), 
\label{ramallotekalzounahostia}
\end{equation}
for some undetermined parameters $a,b,c,d,f,p,q,r,s$, with $q\neq r$ unless $S=0$. However we are not interested in the most general solution, but only in proving that at least one solution exists. Let us show that one solution can be found for the particular choices $c=d=f=0$, with $a,b,p,q,r,s$ to be determined. Having $S=0$ reduces the nonvanishing solutions $P$ of eqn. (\ref{pauli}) to $\pm{\rm i}\sqrt{2}$ times the identity or one of the Pauli matrices $\sigma_x,\sigma_y,\sigma_z$. Neither of the latter leads to an antisymmetric $M$, while for $P=\pm{\rm i}\sqrt{2}{\bf 1}$ we can have $M^t=-M$ if $a=0=b$. Altogether
\begin{equation}
B=\pm{\rm i}\sqrt{2}\left(\begin{array}{cccc}
0&\kappa/4&-1&0\\
-\kappa/4&0&0&-1\\
1&0&0&0\\
0&1&0&0
\end{array}
\right), \qquad \kappa=\frac{e\vert{\bf B}\vert}{c}
\label{javiermasmarikas}
\end{equation}
is one solution to our problem. The overall factor of $\sqrt{-1}$ is irrelevant because it cancels in the transformation law (\ref{ramallokaka}). The ultimate reason for this factor of $\sqrt{-1}$ originates in the conventions chosen in ref. \cite{GUALTIERI} for the basis of $so(2n,2n)$, which we have followed here in eqns. (\ref{barbonketefollenporkulo}), (\ref{ketefollenbarbon}). The inner product (\ref{linprod}) has the (block) matrix representation
\begin{equation}
K=\left(\begin{array}{cc}
0&{\bf 1}_{2n}\\
{\bf 1}_{2n} &0
\end{array}\right).
\label{barbonketedenporkulo}
\end{equation}
The above diagonalises to the metric $G$ of eqn. (\ref{barboneatshit}) as already remarked. Now the unitary matrix $U$
\begin{equation}
U=-\frac{1}{\sqrt{2}}
\left(\begin{array}{cc}
-{\rm i}{\bf 1}_{2n}&{\rm i}{\bf 1}_{2n}\\
{\bf 1}_{2n} &{\bf 1}_{2n}
\end{array}\right)
\label{ramalloketedenporkulo}
\end{equation}
is such that
\begin{equation}
UKU^t={\bf 1}_{4n}.
\label{javiermassodomita}
\end{equation}
Thus the unitary matrix $U^t$ transforms the Euclidean metric on the right--hand side of (\ref{javiermassodomita}) to the split--signature metric $K$ of eqn. (\ref{barbonketedenporkulo}). The latter is equivalent to the metric $G$ of eqn. (\ref{barboneatshit}). In this process, the Lie algebra $so(4n)$ becomes $so(2n,2n)$: this is Weyl's unitary trick. 

Presumably there are more solutions to eqns. (\ref{mekagonramallo}), (\ref{makagoentuputakararamayo}), (\ref{luisibanezmekagontuputakalva}) than given by (\ref{javiermasmarikas}).
Indeed \cite{GUALTIERI}, on a {\it linear}\/ phase space ${\cal C}$, any generalised complex structure of type $k$ can be {\it noncanonically}\/ expressed as a $B$--transform of the direct sum of a complex structure (with complex dimension $k$) and a symplectic structure (of real dimension $2n-2k$). However our solution (\ref{javiermasmarikas}) has the two added bonuses. First, it depends on one real parameter only, the external field $\vert{\bf B}\vert$. Second, the solutions whose existence is guaranteed by the theorems of ref. \cite{GUALTIERI} need not be constant, even in the case when the external magnetic field is uniform. On the contrary our solution is constant by construction, and it allows for a straightforward physical interpretation. Finally, it can also be extended beyond $n=2$ to $n=2d$ degrees of freedom on flat space. On a curved phase space, our constant solution continues to hold locally around any point.

\subsection{Noncommuting momenta {\it and}\/ coordinates}\label{vazquez&ramallo,piojos&liendres}

The Landau problem is usually expressed in terms of noncommuting momenta, while the coordinates remain commutative. By a Fourier transformation we can go over to a dual picture in which the coordinates are noncommuting, 
\begin{equation}
\left\{x,y\right\}=\frac{c}{e\vert {\bf B}\vert},
\label{ramallochupameelcarallo}
\end{equation}
while the momenta are commutative. An instance in which neither the momenta Poisson--commute among themselves, nor the coordinates, while not exactly corresponding to the Landau problem, has been analysed recently in ref. \cite{BRAZIL} using symplectic techniques. By an appropriate choice of the $B$--field, generalised complex manifolds also allow to express a Poisson structure with noncommuting coordinates and noncommuting momenta such as
\begin{equation}
\left\{x,y\right\}=\frac{c}{e\vert {\bf B}\vert}, \quad \left\{p_x,p_y\right\}=\frac{e \vert{\bf B}\vert}{c}, \quad \left\{x,p_x\right\}=1=\left\{y,p_y\right\},
\label{bazz}
\end{equation}
in terms of the $B$--transform of the canonical Poisson brackets (\ref{barbonketemetanuncilindro}). Hence the range of applicability of our technique is not limited to the Landau problem.

\section{Discussion}\label{javiermasimbecildelculo}

The interaction of a charged particle with an external magnetic field is implemeted in symplectic geometry by adding a suitable {\it magnetic term}, linear in the magnetic field, to the symplectic form  \cite{MR}. Such redefinitions have been known for long. However this procedure is purely {\it ad hoc}: its justification lies in the fact that it transforms the Poisson brackets $\left\{p_x,p_y\right\}=0$ into the new brackets $\left\{p_x,p_y\right\}=e\vert{\bf B}\vert/c$. Generalised complex geometry presents the advantage that, interpreting this redefinition as a $B$--transformation, the magnetic term of the symplectic form appears naturally, without having to be put in by hand. In turn, $B$--transformations are a consequence of the split--signature metric that every generalised complex manifold carries, and its invariance group $SO(2n,2n)$. In this way the key issue becomes the following: what is the physical origin for a metric of signature $(2n,2n)$ on phase space? Once this has been accounted for, the $B$--field and the corresponding $B$--transformations follow naturally. This question appears puzzling on first sight since, in the presence of $n$ degrees of freedom, phase space ${\cal C}$ is $2n$--dimensional. The answer lies in simultaneously placing all $2n$ classical coordinates $q^j,p_j$, plus their quantum counterparts, on the same footing. This is a central idea in the theory of generalised complex manifolds, where the object of study is $T{\cal C}\oplus T^*{\cal C}$ rather than $T{\cal C}$ or $T^*{\cal C}$ alone.

In section \ref{javiermasmariquitas} we have proved that there is a simple origin for the split--signature metric in terms of the classical Poisson brackets and their quantum counterparts. Namely, the matrix representing classical Poisson brackets in Darboux coordinates squares to minus the identity. Multiplication by $\sqrt{-1}$ gives quantum Poisson brackets, {\it i.e.}, commutators. The factor of $\sqrt{-1}$ ensures that the matrix representing the latter in Darboux coordinates squares to the identity. Altogether this gives a physical origin for the split--signature metric of generalised complex manifolds. The same setup arose in ref. \cite{ME} in connection with duality transformations between {\it classical}\/ and {\it quantum}. Our main results follow from the transformation law (\ref{ramallokaka}) for the symplectic form under a $B$--field. We have shown that it converts the standard Poisson brackets $\left\{p_x,p_y\right\}=0$ into the new brackets $\left\{p_x,p_y\right\}=e\vert{\bf B}\vert/c$, as in the Landau problem. Our conclusions can be summarised in the statement that generalised complex geometry provides a way to study physics in the presence of a magnetic background.

In section \ref{ramalloguarrolavatelospies} we have argued that the geometry underlying the $B$--field escapes Klein's {\it Erlanger Programm}, in the sense that it cannot be characterised simply as being invariant under some group of transformations. While there is certainly an isometry group $SO(2n,2n)$ acting on phase space ${\cal C}$, a key role is played by the $B$--transformation law for the symplectic form, eqn. (\ref{ramallokaka}). Since $B$--transformations are not diffeomorphisms of ${\cal C}$, this explains our statement.

{\bf Acknowledgements}

It is a great pleasure to thank J. de Azc\'arraga for encouragement and support and U. Lindstr\"om for drawing attention to some references. The author thanks Max-Planck-Institut f\"ur Gravitationsphysik (Potsdam, Germany) where this work was begun, for hospitality. This work has been partially supported by EU network MRTN--CT--2004--005104, by research grant BFM2002--03681 from Ministerio de Ciencia y Tecnolog\'{\i}a, by research grant GV2004--B--226 from Generalitat Valenciana, by EU FEDER funds, by Fundaci\'on Marina Bueno and by Deutsche Forschungsgemeinschaft.

\end{document}